\documentclass[11pt]{article}
\usepackage{longtable,supertabular}
\usepackage{amsmath}
\usepackage{amssymb}
\usepackage{latexsym}
\usepackage{cite}

\title{\bf Storage and Retrieval Codes in PIR Schemes with Colluding Servers}
\author{Hao Chen and Liqing Xu
  \thanks{Hao Chen and Liqing Xu are with the College of Information Science and Technology/Cyber Security, Jinan University, Guangzhou, Guangdong Province, 510632, China, haochen@jnu.edu.cn, lqxu1@jnu.edu.cn. The research was supported by NSFC Grant 62032009.}}

\begin{document}

\maketitle
\begin{abstract}
Private information retrieval (PIR) schemes (with or without colluding servers) have been proposed for realistic coded distributed data storage systems. Star product PIR schemes with colluding servers for general coded distributed storage system were constructed over general finite fields by R. Freij-Hollanti, O. W. Gnilke, C. Hollanti and A. Karpuk in 2017. These star product PIR schemes with colluding servers are suitable for the storage of files over small fields and can be constructed for coded distributed storage system with large number of servers. In this paper for an efficient storage code, the problem to find good retrieval codes is considered. In general if the storage code is a binary Reed-Muller code the retrieval code needs not to be a binary Reed-Muller code in general. It is proved that when the storage code contains some special codewords, nonzero retrieval rate star product PIR schemes with colluding servers can only protect against small number of colluding servers. We also give examples to show that when the storage code is a good cyclic code, the best choice of the retrieval code is not cyclic in general. Therefore in the design of star product PIR schemes with colluding servers, the scheme with the storage code and the retrieval code in the same family of algebraic codes is not always efficient.\\

\end{abstract}

\section{Introduction}

Distributed data storage from a linear code ${\bf C} \subset {\bf F}_q^n$ can be used to protect information when some servers are failed. The private information retrieval for coded distributed storage system from a general linear code ${\bf C} \subset {\bf F}_{q^b}^n$ was considered in \cite{Freij,Freij1,Taje}.  $m$ files ${\bf x}_1, \ldots, {\bf x}_m \in {\bf F}_q^{b \times k}$ need to be stored and downloaded across $n$ servers. Each ${\bf x}_i=(x_1^i, \ldots, x_k^i)$ can be considered as a length $k$ vector in ${\bf F}_{q^b}$ for $i=1, \ldots, m$. The files can be considered as a $m \times k$ matrix ${\bf X}=(x_j^i)_{1 \leq i \leq m, 1 \leq j \leq k}$ with entries in ${\bf F}_{q^b}$. Let ${\bf G}({\bf C})$ be a generator matrix of the linear code ${\bf C}$ with $n$ columns ${\bf g}_1, \ldots, {\bf g}_n \in {\bf F}_{q^b}^k$. Then a length $m$ vector $(<{\bf x}_1, {\bf g}_i>, \ldots, <{\bf x}_m, {\bf g}_i>) \in {\bf F}_{q^b}^m$ is stored in the $i$-th server. When the files are downloaded from servers, then if at most $d({\bf C})-1$ servers are failed, then files can be recovered. Actually among any $n-d({\bf C})+1$ columns in ${\bf g}_1, \ldots, {\bf g}_n$, there are $k$ linear independent columns, otherwise there is a weight $d({\bf C})-1$ codeword in ${\bf C}$. We call $f=\frac{d({\bf C})-1}{n}$ the ratio of tolerated failed servers of this coded distributed storage system. The storage rate of this distributed storage system is $R_{storage}=\frac{bk}{nb}=\frac{k}{n}$. This code is called the storage code. When ${\bf C}$ is the $[n, 1, n]_{q^b}$ code,  this is the replicated data storage. From the Singleton bound $$R_{storage}+f \leq 1.$$  Hence if the ratio of tolerated failed servers is high the storage rate is low. When ${\bf C}$ is an MDS $[n, k, n-k+1]_q$ code, this is the MDS coded distributed data storage system. Any $k$ servers can recover the files if at most $n-k$ servers are failed. In MDS-coded distributed storage system the storage code is an BCH code, see Section 6 below. If the storage code ${\bf C}$ is not MDS, then at least $n-d+1\geq k+1$ servers are needed to recover the files. \\

Private information retrieval was proposed and studied in 1995 by Chor, Goldreich, Kushilevitz and Sudan, see \cite{Chor,Chor1}. In the classical model data ${\bf x}=(x^1, \ldots, x^m) \in {\bf F}_2^m$ is stored and the user want to retrieval a single bit $x^i$ without revealing any information about the index $i$. The motivation of the construction of PIR scheme is to reduce the total download complexity, see \cite{Beimel}. The retrieval rate of an PIR scheme is the ratio of the gained information over downloaded information. PIR scheme for more realistic coded distributed data storage system was first studied in \cite{Shah,Blackburn}. All files ${\bf x}_1, \ldots, {\bf x}_m \in {\bf F}_q^{b \times k}$ are stored over servers according to a storage code ${\bf C}$ as above. The coded distributed data storage can be used to help to recover the data in the case of server failure. PIR scheme with $t$ colluding servers is a protocol in which the user want to retrieval the  file ${\bf x}_w$ without revealing any information of the index $w$, while any $\leq t$ servers can share their quaries from the user.  PIR schemes for MDS coded distributed data (or replicated data) storage were constructed in previous papers \cite{Banawan,Freij,SunJafar,SunJafar1} and references therein. The MDS coded distributed storage system is maximally robust against server failure. For these coded distributed storage, the files are in ${\bf F}_q^{b \times k}$ with the condition $q^b \geq n$. That is, each file has at least $log_2 n$ bits. Because MDS codes over ${\bf F}_{q^b}$ satisfying $q^b\geq n$ are used, the download complexity of the user and the computational complexity of each server are lower bounded by a $log_2n$ factor, since a large field has to be used. These schemes are not so suitable when files with bounded length of bits are stored across large number of servers. PIR schemes with colluding servers for coded databases from general linear codes over small fields are considered in \cite{Freij,Freij1,Lin,Kumar,KLRA}. We refer to \cite{BL20,Fazeli,HHW,LN21,Tian,LHK,LKRAY,Wang,Taje1} for the recent related development and \cite{Ulukus} for a nice survey of active research on private retrieval, computing and learning.\\

Star product PIR schemes with $t$-colluding servers for coded distributed storage from the storage code ${\bf C} \subset {\bf F}_q^n$ was constructed from another retrieval code ${\bf D} \subset {\bf F}_q^n$ in \cite{Freij,Freij1,Lin} . The star product PIR scheme has the retrieval rate $$R_{retrieval}=\frac{d({\bf C} \star {\bf D})-1}{n}$$ for general linear codes ${\bf C}$ and ${\bf D}$, see Theorem 2 in page 2111of  \cite{Freij1}. When both ${\bf C}$ and ${\bf C} \star {\bf D}$ are transitive the rate can be improved to $$R_{retrieval}=\frac{\dim(({\bf C} \star {\bf D})^{\perp})}{n},$$ see Theorem 4 in page 2113 of \cite{Freij1}.  It is clear that $$d({\bf C} \star {\bf D}) -1 \leq \dim(({\bf C} \star {\bf D})^{\perp})$$ from the Singleton bound for the linear code ${\bf C} \star {\bf D}$. Their PIR schemes protect against $d({\bf D}^{\perp})-1$ colluding servers. This is called a $(t=d({\bf D}^{\perp})-1)$-privacy PIR scheme. \\

Reed-Muller codes are length $2^m$ binary codes were introduced independently in 1954, see \cite{Muller,Reed}. The Reed-Muller code $RM(m, r)$ is a linear $[2^m, \Sigma_{i=0}^r \displaystyle{m \choose i}, 2^{m-r}]_2$ code. The dual code is $RM(m, m-r-1)$. Punctured Reed-Muller codes can be considered as length $2^m-1$ cyclic codes, see \cite{BS17,DLX}. Reed-Muller code based star product PIR scheme with colluding servers were studied in \cite{Freij1,Saarela,LN21}. These $t$-privacy PIR schemes for coded database can be defined over the smallest field ${\bf F}_2$ and have good performance. The star product PIR scheme with colluding servers from the generalized Reed-Solomon codes were studied in \cite{Freij}. These $t$-privacy PIR for MDS-coded database can only be defined over the field ${\bf F}_q^{b \times k}$ satisfying $n \leq q^b$. Star product private information retrieval schemes using cyclic codes for $2^m-1$ server distributed storage were studied in \cite{BMR21}. The parameters were calculated and compared with parameters of punctured Reed-Muller code (a special cyclic code of length $2^m-1$ based star product PIR schemes. It was proved that the $2^m-1$-server cyclic coded star product PIR schemes outperform punctured RM code based star product PIR schemes.\\

First of all we need an efficient storage code ${\bf C} \subset {\bf F}_q^n$ with $k+d$ as large as possible. For example when $n=64$ servers are used and at most $16-1$ servers can be failed when files are reading from these servers. The storage rate is $\frac{27}{64}$ when ${\bf C}$ is a linear $[64, 27, 16]_2$ code. This is from the shortening $[64, 27, 16]_2$ code ${\bf C}\subset {\bf F}_2^{64}$ of the $[73, 36, 16]_2$ cyclic code, see \cite{codetable}. The storage rate of $RM(6, 2)$-coded distributed storage is $\frac{22}{64}<\frac{27}{66}$.  Therefore to construct efficient star product PIR scheme it is more reasonable to find a good retrieval code ${\bf D}$  for this $[64, 27, 16]_2$ shortening storage code ${\bf C}$ of the cyclic code.  On the other hand there is a binary linear $[64, 28, 16]_2$ code, see \cite{codetable}. It is difficult to calculate the star product for this code and a retrieval code to construct a star product PIR scheme with colluding servers for this $[64, 28, 16]_2$ coded distributed storage system. \\

When an efficient storage code ${\bf C}$ is fixed, the problem to find a retrieval code ${\bf D}$ such that the star product PIR scheme constructed in \cite{Freij,Freij1} has large retrieval rate $\frac{d({\bf C} \star {\bf D})-1}{n}$ (or $\frac{\dim(({\bf C} \star {\bf D})^{\perp})}{n}$ when both ${\bf C}$ and ${\bf C} \star {\bf D}$ are transitive) and large privacy number $d({\bf D}^{\perp})-1$, is interesting. Thus we have the following constructive problem of linear storage and retrieval codes about star product PIR schemes with colluding servers.\\

{\bf Storage and Retrieval Code Problem.} {\em To find two linear codes ${\bf C} \subset {\bf F}_q^n$ and ${\bf D} \subset {\bf F}_q^n$ such that\\
1) The defect of the code ${\bf C}$, $D({\bf C})=n-(d({\bf C})+\dim({\bf C}))$ is small to make the distributed storage system efficient;\\
2) $d({\bf C} \star {\bf D})-1$ is as large as possible; or $\dim({\bf C} \star {\bf D})$ is as small as possible when ${\bf C}$ and ${\bf C} \star {\bf D}$ are transitive codes;\\
3) $d({\bf D}^{\perp})$ is as large as possible.}\\

The Singleton bound in $$\dim(({\bf C} \star {\bf D})^{\perp})+\dim({\bf C})+d({\bf D}^{\perp}) \leq n+2$$ was proved for star product PIR schemes with colluding serves in \cite{Chen}. We call the difference $(n+2)-\dim(({\bf C} \star {\bf D})^{\perp})+\dim({\bf C})+d({\bf D}^{\perp})$ the defect of this star product PIR schemes with colluding servers and denoted by $$D(PIR({\bf C} \star {\bf D}))=(n+2)-\dim(({\bf C} \star {\bf D})^{\perp})+\dim({\bf C})+d({\bf D}^{\perp}).$$

 In this paper the problem to find good retrieval code ${\bf D}$ for a good storage code ${\bf C}$ such that the retrieval rate and the privacy $t$ are as large as possible is considered. This proposes an interesting problem for coding theory which are related to the star product of two linear codes and the dual code of the second code. In the simplest case of replicated data storage that ${\bf C}$ is a linear $[n, 1, n]_q$ code, we illustrate that the cyclic or BCH retrieval code is a good choice for small $q$, and the algebraic geometric retrieval code is a good choice when $q\geq 5$. Many known good binary linear codes are punctured codes or shortening codes from BCH codes or extended BCH code, see \cite{codetable}. Therefore it seems reasonable to consider the above problem for good cyclic storage code. In Section 3 and 4 examples are given to show that when the storage code is a good cyclic code, the best choice of the retrieval code is not always cyclic in general. In Section 5, it is proved that when the storage code contains some special codewords, nonzero retrieval rate star product PIR schemes can only protect against small number of colluding servers. In Section 6 the retrieval rates of cyclic code based star product PIR schemes with colliding servers are lower bounded. Low cost cyclic code based star product PIR schemes with colluding servers are constructed from cyclic storage code with two nonzero weights studied in the classical paper \cite{BM72}. In Section 7 we show that when the storage code ${\bf C}$ is the Reed-Muller code which is an extended cyclic code, in general the best choice of the retrieval code is not the Reed-Muller code as in \cite{Freij1,Saarela}.\\

\section{Preliminaries}

The Hamming weight $wt({\bf a})$ of a vector ${\bf a} \in {\bf F}_q^n$ is the number of non-zero coordinate positions. The Hamming distance between two vectors ${\bf a}$ and ${\bf b}$ is the Hamming weight of ${\bf a}-{\bf b}$, $d({\bf a}, {\bf b})=wt({\bf a}-{\bf b})$. The Hamming distance of a code ${\bf C} \subset {\bf F}_q^n$ is $$d({\bf C})=\min_{{\bf a} \neq {\bf b}} \{d({\bf a}, {\bf b}),  {\bf a} \in {\bf C}, {\bf b} \in {\bf C} \}.$$  The minimum Hamming distance of  a linear code is its minimum Hamming weight.  For a linear $[n, k, d]_q$ code, the Singleton bound asserts $d \leq n-k+1$. When the equality holds, this code is an MDS code. Reed-Solomon codes introduced in \cite{RS} are well-known MDS codes. We refer to \cite{HP} for the theory of Hamming error-correcting codes. Let $\sigma \in {\bf S}_n$ be a permutation of $n$ coordinate positions, where ${\bf S}_n$ is the order $n!$ group of all permutations of $n$ symbols, if $\sigma({\bf C})={\bf C}$, $\sigma$ is an automorphism of this code. The subgroup of all automorphisms of this code is denoted by $Aut({\bf C})$. The code is called transitive if $Aut({\bf C})$ acts transitively at the set of $n$ symbols, that is, if for any two coordinate position $u, v \in [n]$, there is an automorphism $\sigma$ of ${\bf C}$ such that $\sigma(u)=v$.\\

The dual of a linear code ${\bf C}\subset {\bf F}_q^n$ is $${\bf C}^{\perp}=\{{\bf c} \in {\bf F}_q^n: <{\bf c}, {\bf y}>=\Sigma_{i=1}^n c_i y_i=0, \forall {\bf y} \in {\bf C}\}.$$  The minimum distance of the Euclidean dual is called the dual distance and is denoted by $d^{\perp}$. The componentwise product (star product) of $t$ vectors ${\bf x}_j=(x_{j,1}, \ldots, x_{j, n}) \in {\bf F}_q^n$, $j=1, \ldots, t$, is ${\bf x}_1 \star \cdots \star {\bf x}_t=(x_{1,1} \cdots x_{t,1}, \ldots, x_{1, n} \cdots x_{t,n}) \in {\bf F}_q^n$. The componentwise product (star product) of linear codes ${\bf C}_1, \ldots, {\bf C}_t$ in ${\bf F}_q^n$ is defined by $${\bf C}_1\star \cdots \star {\bf C}_t=\Sigma_{ {\bf c}_i \in {\bf C}_i} {\bf F}_q {\bf c}_1 \star \cdots \star {\bf c}_t, $$ we refer to \cite{Randri,Cascudo} for the study of this product.\\

Let ${\bf F}_q$ be a finite field. Then a linear code ${\bf C} \subset {\bf F}_q^n$ is called cyclic if $(c_0, c_1, \ldots, c_{n-1}) \in {\bf C}$, then $(c_{n-1}, c_0, \ldots, c_{n-2}) \in {\bf C}$. A codeword ${\bf c}$ in a cyclic code is identified with a polynomial ${\bf c}(x)=c_0+c_1x+\cdots+c_{n-1}x^{n-1}\in {\bf F}_q[x]/(x^n-1)$. Every cyclic code is generated by a factor polynomial of $x^n-1$. The dimension of the cyclic code ${\bf C}_{{\bf g}}$ generated by ${\bf g}(x)=g_0+g_1x+\cdots+g_{n-k}x^{n-k} \in {\bf F}_q[x]$ is $n-\deg({\bf g}(x))=k$. The dual of a cyclic code is a cyclic code generated by ${\bf g}^{\perp}=\frac{x^k{\bf h}(x^{-1})}{{\bf h}(0)}$, where ${\bf h}(x)=\frac{x^n-1}{{\bf g}(x)}$, see \cite{HP} Chapter 4.\\

We assume $\gcd(n, q)=1$. Let $m$ be the smallest positive integer satisfying $q^m \equiv 1$ $mod$ $n$. Let $\alpha$ be a primitive element of the extension field ${\bf F}_{q^m}$. Then $\beta=\alpha^{\frac{q^m-1}{n}}$ is a primitive $n$-th root of unity. Each root of $x^n-1$ is of the form $\beta^i$, $i=0, 1, \ldots, n-1$. Set ${\bf Z}_n={\bf Z}/n{\bf Z}=\{0, 1, \ldots, n-1\}$. A subset ${\bf T} \subset {\bf Z}_n$ is called a cyclotomic coset if $${\bf C}_i=\{i, iq, \ldots, iq^{l-1}\},$$ where $i \in {\bf Z}_n$ is fixed and $l$ is the smallest positive integer such that $iq^l \equiv i$ $mod$ $n$. It is clear that cyclotomic cosets corresponds to irreducible factors in ${\bf F}_q[x]$ of of $x^n-1$. Therefore a generator polynomial of a cyclic code is the product of several irreducible factors of $x^n-1$. The defining set of a cyclic code generated by ${\bf g}$  is the the following set $${\bf T}_{{\bf g}}=\{i: {\bf g}(\beta^i)=0\}.$$ The defining set of a cyclic code is the disjoint union of several cyclotomic cosets. Set $${\bf T}^{-1}=\{-i: i \in {\bf T}\},$$  Then the defining set of the dual code is $({\bf T}^c)^{-1}$, see \cite{HP} Chapter 8.\\

Let ${\bf m}_i(x)$ be the minimal polynomial of $\beta^i$ over ${\bf F}_q$. For any designed distance $\delta \geq 2$, we set $${\bf g}_{q, n, \delta, b}(x)=lcm({\bf m}_b(x), {\bf m}_{b+1}(x), \ldots, {\bf m}_{b+\delta-2}(x)),$$ where $lcm$ is the least common multiple of polynomials in ${\bf F}_q[x]$. The cyclic code ${\bf C}_{q, n, \delta, b}$ generated by ${\bf g}_{q, n, b, \delta}(x)$ is called Bose-Chaudhuri-Hocquenghem (BCH) code introduced in \cite{BC1,BC2,Hoc}. It was proved in \cite{BC1,BC2,Hoc} the minimum Hamming distance is larger than or equal to the designed distance. When $n=p^m-1$ the BCH code is called primitive and when $b=1$ the BCH code is called narrow-sense. The BCH codes have been studied for many years and served as basic examples of good linear error-correcting codes. Parameters of many BCH codes have been determined in \cite{LDL,DLX}. \\

Let ${\bf F}_q$ be an arbitrary finite field, $P_1,\ldots,P_n$ be $n \leq q$ elements in ${\bf F}_q$. The Reed-Solomon code  $RS(n,k)$ is defined by $$RS(n,k)=\{(f(P_1),\ldots,f(P_n)): f \in {\bf F}_q[x],\deg(f) \leq k-1\}.$$ This is an MDS $[n,k,n-k+1]_q$ linear code attaining the Singleton bound $d \leq n-k+1$ since  a degree $\deg(f) \leq k-1$ nonzero polynomial has at most $k-1$ roots. PIR schemes studied in \cite{Banawan,SunJafar,SunJafar1,Freij,Wang} are for MDS coded databases. They are based on the generalized Reed-Solomon codes.\\

Let ${\bf X}$ be an absolutely irreducible, smooth and genus $g$ curve defined over ${\bf F}_q$. Let $P_1,\ldots,P_n$ be $n$ distinct rational points of ${\bf X}$ over ${\bf F}_q$. Let ${\bf G}$ be a rational divisor over ${\bf F}_q$ of degree $\deg({\bf G})$ satisfying $2g-2 <\deg({\bf G})<n$ and $$support({\bf G}) \bigcap {\bf P}=\emptyset.$$ Let ${\bf L}({\bf G})$ be the function space associated with the divisor ${\bf G}$. The algebraic geometry code (functional code) associated with ${\bf G}$, $P_1,\ldots,P_n$ is defined by $${\bf C}(P_1,\ldots,P_n, {\bf G}, {\bf X})=\{(f(P_1),\ldots,f(P_n)): f \in {\bf L}({\bf G})\}.$$ The dimension of this code is $$k=\deg({\bf G})-g+1$$ follows from the Riemann-Roch Theorem. The minimum Hamming distance is $$d \geq n-\deg({\bf G}).$$ An divisor ${\bf G}=\Sigma m_i G_i$ where $m_i \geq 0$ is called an effective divisor. Algebraic-geometric residual code ${\bf C}_{\Omega}(P_1, \ldots, P_n, {\bf G}, {\bf X})$ with the dimension $k=n-m+g-1$ and minimum Hamming distance $d \geq m-2g+2$ can be defined, we refer to \cite{HP,TV} for the detail. It is the dual code of the functional code of the dimension $m-g+1$.\\

Let ${\bf E}$ be an elliptic curve defined over ${\bf F}_q$.  It is well-known that when $q$ is not a power of $2$ or $3$, then elliptic curves over ${\bf F}_q$ can be realized as a non-singular plane cubic curve. Let ${\bf E}({\bf F}_q)$ be the set of all ${\bf F}_q$-rational points of ${\bf E}$. This is an Abelian group, for its group structure, we refer to \cite{Ruck}. The number $|{\bf E}({\bf F}_q)|$ of its rational points over ${\bf F}_q$ satisfies the Hasse bound $$|q+1-|{\bf E}({\bf F}_q)|| \leq 2\sqrt{q}.$$  For any positive real number $x$ we set $$x^{-}=x+1-2\sqrt{x}, $$ and $$x^{+}=x=1+2\sqrt{x}.$$ If $q=p$ is a prime number it follows from the result in \cite{Deuring,Ruck} that for any positive integer $N$ satisfying $p^{-} <N < p^{+}$, there is an elliptic curve ${\bf E}$ defined over ${F}_p$ such that the number of ${\bf F}_p$-rational points of ${\bf E}$ satisfying $$|{\bf E}({\bf F}_p)|=N.$$

\section{Retrieval codes for the replicated data storage}

If the storage code is $[n, 1, n]_q$ replicate code. Let the retrieval code be ${\bf D} \subset {\bf F}_q^n$. Then the star product PIR scheme introduced in \cite{Freij1} protects against $d({\bf D}^{\perp})-1$ colluding servers and has the retrieval rate at least $\frac{d({\bf D})-1}{n}$. Then we need to find a linear code ${\bf D} \subset {\bf F}_q^n$ such that $d({\bf D})+d({\bf D}^{\perp})$ as large as possible. From the Singleton bound $$d({\bf D}^{\perp})\leq \dim({\bf D})+1,$$ then $$d({\bf D})+d({\bf D}^{\perp}) \leq \dim({\bf D})+d({\bf D})+1=n+2.$$ The equality hold if and only if ${\bf D}$ is an MDS code. However when the number $n$ of servers is much bigger than $q$, there is no MDS code with the length $n$.\\

In general we need to find a linear transitive retrieval code ${\bf D}$ such that $\dim({\bf D}^{\perp})+d({\bf D}^{\perp})$ as large as possible When ${\bf D}$ is a cyclic code and ${\bf D}^{\perp}$ is an BCH code we have the lower bound on its dimension $$\dim({\bf C}_{q, q^m-1, b, \delta}) \geq q^m-1-(\delta-1)m$$ for $q^m-1$ server distributed storage system. Then $$\dim({\bf D}^{\perp})+d({\bf D}^{\perp}) \geq n-(\delta-1)m+\delta.$$ On the other hand algebraic geometric codes are suitable choices. Since $d({\bf D}^{\perp})+d({\bf D}) \geq n-2g+2$ for any algebraic geometric code ${\bf D}$.  For the replicate data storage, when the base field is binary or a small field, the cyclic or BCH retrieval code is a good choice. When the field is of the size $q\geq 5$, the choice of AG retrieval code is better.\\

{\bf Example 3.1.} We consider the $\frac{2^9-1}{7}=73$-server replicated data storage over the binary field. If we use the retrieval code ${\bf D}$ as the dual of the $[73, 36, 16]_2$ cyclic code with the generator polynomial $x^{37}+x^{36}+x^{34}+x^{33}+x^{32}+x^{27}+x^{25}+x^{24}+x^{22}+x^{21}+x^{19}+x^{18}+x^{15}+x^{11}+x^{10}+x^8+x^7+x^5+x^3+1$, see \cite{codetable}. Then the retrieval rate is $\frac{\dim({\bf D}^{\perp})}{73}=\frac{36}{73}$ and the scheme protects against $15$ colluding servers.\\

{\bf Example 3.2.} We consider the $73$-server replicated data storage over the field ${\bf F}_{16}$. There is a genus $10$ curve defined over ${\bf F}_{16}$ with at least $81$ rational points. Then we can use the dual code of an AG $[73, 48, 16]_{2^{16}}$ code as the retrieval code in the star product PIR scheme with colluding servers. The dual codes is an AG $[73, 25, 39]_{2^{16}}$ code. The retrieval rate is at least $\frac{39-1}{73}=\frac{38}{73}$ and the scheme protects against $15$ colluding servers.\\

{\bf Example 3.3.} We consider the $7$-server replicated data storage over ${\bf F}_5$. The storage code is the trivial cyclic $[7, 1, 7]_5$ cyclic code. If the retrieval code is restricted to cyclic code of the length $7$ over ${\bf F}_5$. Since $x^7-1=(x-1)(x^6+x^5+x^4+x^3+x^2+x+1)$ is the decomposition of $x^7-1 \in {\bf F}_5[x]$ to irreducible factors. This is obvious from the factor the cyclotomic cosets of ${\bf Z}/7{\bf Z}$ has only two cyclotomic cosets $C_0=\{0\}$ and $C_1=\{1, 2, 3, 4, 5, 6\}$, since $5^6\equiv 1$ $mod$ $7$ and $5$ is a primitive root of $1$ module $7$. The two choices of cyclic retrieval code are $[7, 1, 7]_5$ code and $[7, 6, 2]_5$ code. The corresponding star product PIR scheme with colluding servers can protect $7-1=6$ colluding servers with the retrieval rate $\frac{1}{7}$. The another star product PIR scheme with colluding servers can protect to $2-1=1$ colluding servers with the retrieval rate $\frac{6}{7}$. If there are $2$ colluding serves in the replicated data storage system. Only the first PIR scheme with the small retrieval rate $\frac{1}{7}$ can be used.\\

On the other hand from the classical result about elliptic curves, see \cite{Ruck}, there is an elliptic curve defined over ${\bf F}_5$ with $8$ rational points. We have an AG $[7, m, 7-m]_5$ code ${\bf D}$ with the ${\bf D}^{\perp}$ is an elliptic curve AG $[7, 7-m, m]_5$ code, where $1\leq m \leq 6$. Therefore we have AG code based star product PIR scheme with colluding servers protect against to $m-1$ colluding servers with the retrieval rate $\frac{d({\bf D})-1}{7}=\frac{6-m}{7}$. When $m=3$ we have an elliptic curve AG code based star product PIR scheme protects against $2$ colluding servers with the retrieval rate $\frac{4}{7}$. This example show that when the number of servers are in the rage $p+2 \leq n \leq p+1+2\lfloor \sqrt{p} \rfloor$. AG code based star product PIR scheme with colluding servers for replicated data storage is better than cyclic code based star product PIR scheme.\\

\section{AG retrieval codes for cyclic storage codes}

Let ${\bf C}$ be a linear cyclic $[2n, 2, n]_q$ code with the following generator matrix\\

$$
\left(
\begin{array}{cccccccccccccccc}
1&0&1&0&\cdots&\cdots&1&0\\
0&1&0&1&\cdots&\cdots&0&1\\
\end{array}
\right)
$$

Then for any retrieval code ${\bf D} \subset {\bf F}_q^{2n}$, let ${\bf D}_1 \subset {\bf F}_q^{2n}$ be the punctured code at odd coordinate positions and ${\bf D}_2 \subset {\bf F}_q^{2n}$ be the punctured code at even coordinate positions. We have ${\bf C} \star {\bf D}={\bf D}_1 \oplus {\bf D}_2$, where zero coordinates are padded at punctured positions. When $n$ is some positive integer such that there are very few cyclic codes of the length $n$, the retrieval cyclic code is not a good choice. For example when $q=5$ and $n=7$. There are only $9$ choice of such cyclic codes with $d^{\perp}=1$ (no colluding server at all), $d^{\perp}=2$ (only one colluding server allowed), $d^{\perp}=6$ (retrieval rate $\frac{2}{14}=\frac{1}{7}$), and $d^{\perp}=14$. The last one star product PIR scheme has the zero retrieval rate. If there are $2$ colluding severs we have to use the star product PIR scheme with the very small retrieval rate $\frac{1}{7}$, if only cyclic retrieval code is allowed.\\

On the other hand, we can use the direct sum of two elliptic curve AG $[7, 3, 4]_5$ codes as in Example 2.3 as the retrieval code. The dual code is the direct sum of two elliptic curve $[7, 4, 3]_5$ code, which is a $[14, 8, 3]_5$ code. This star product PIR scheme protects against $2$ colluding servers with the retrieval rate $\frac{4-1}{14}=\frac{3}{14}$.\\

It is obvious that the above example can be generalized to the case there are very few length $n$ cyclic code case. For example when $q=2p+1$, $p$ prime, then ${\bf F}_q$ has primitive elements with the order $2p$. Then any prime in the range $1\leq p' \leq q-1$ has the order $2p, p, 2$. There are very few length $q$ cyclic codes over ${\bf F}_{p'}$. Therefore we show that for many cases even if the storage code is a good cyclic code, the best choice of the retrieval code is not the cyclic code.\\

\section{The upper bound on the number of colluding servers}

In this section we show that when the storage code contains some special codewords, then the number of colluding servers in star product PIR schemes is restricted no matter how to choose the retrieval code.\\

{\bf Theorem 5.1.} {\em Suppose that the linear storage code ${\bf C}_1 \subset {\bf F}_q^n$ contains $t$ codewords of the weights $w_1, w_2, \ldots, w_t$ satisfying $$w_1+w_2+\cdots+w_t=n$$ and disjoint supports. Then a nonzero retrieval rate star product PIR scheme with colluding servers can not protect against more than $$\max \{w_1, w_2, \ldots, w_t\}-1$$ colluding servers.}\\

{\bf Proof.} Without loss of generality the two codewords are of the form ${\bf c}_1=(y_1, \ldots, y_{w_1}, 0, \ldots, 0), \ldots, {\bf c}_t=(0, \ldots, 0, y_{w_1+\cdots+w_{t-1}+1}, \ldots, y_n)$. Let ${\bf D}$ be a retrieval code. Then ${\bf D} \subset {\bf D}_1 \oplus {\bf D}_2 \oplus \cdots \oplus {\bf D}_t$ where $${\bf D}_1=\{(c_1, c_2, \ldots, c_{w_1}): (c_1, \ldots, c_{w_1}, {\bf x}) \in {\bf D},  \exists {\bf x} \in {\bf F}_q^{w_2+\cdot+w_t}\},$$ $${\bf D}_2=\{(c_1, c_2, \ldots, c_{w_2}): ({\bf y}, c_1, \ldots, c_{w_2}) \in {\bf D}, \exists {\bf y} \in {\bf F}_q^{w_1+w_3+\cdots+w_t}\},$$ $$......,$$  $${\bf D}_t=\{(c_1, c_2, \ldots, c_{w_t}): ({\bf y}, c_1, \ldots, c_{w_t}) \in {\bf D}, \exists {\bf y} \in {\bf F}_q^{w_1+\cdots+w_{t-1}}\}$$ It is clear ${\bf D}_1^{\perp} \oplus {\bf D}_2^{\perp}\oplus \cdots {\bf D}_t^{\perp} \subset {\bf D}^{\perp}$. For a Hamming weight $w_i$ vector ${\bf v}=(v_1, v_2, \ldots, v_{w_i}) \in {\bf F}_q^{w_i}$ and a code ${\bf E} \subset {\bf F}_q^{w_i}$, set ${\bf v} \cdot {\bf E}=\{(v_1c_1, \ldots, v_{w_i}c_{w_i}): \forall {\bf c}=(c_1, \ldots, c_{w_i}) \in {\bf E}\}$. Since the retrieval rate of this PIR scheme with colluding servers is nonzero, then $\dim({\bf D}_1)+\dim ({\bf D}_2)+\cdots+\dim({\bf D}_t) <n$, otherwise ${\bf c}_1 \cdot {\bf D}_1 \oplus {\bf c}_2 \cdot {\bf D}_2 \oplus \cdots \oplus {\bf c}_t \cdot {\bf D}_t={\bf C} \star {\bf D}$. Therefore ${\bf D}_1^{\perp} \oplus {\bf D}_2^{\perp} \oplus \cdots \oplus {\bf D}_t^{\perp}$ is not zero. Then $$d({\bf D}^{\perp}) \leq \max \{w_1, w_2, \ldots, w_t\}.$$ The conclusion is proved.\\

The above result gives a strong upper bound on the number of colluding servers in the star product PIR scheme with colluding servers. For example when the storage code is the $[8, 4, 4]_2$ Hamming code, the number of colluding servers is at most $3$. When the storage code is the binary Golay $[24, 12, 8]_2$ code, the number of colluding servers is at most $7$. When the storage code is the binary quadratic residue code,  the number of colluding servers in the star product PIR scheme can not be more than $\frac{n}{2}$. When the storage code is the ternary Golay $[12, 6, 6]_3$ code, the number of colluding servers in the star product PIR scheme can not be more than $5$.\\

{\bf Corollary 5.1.} {\em Suppose that the linear storage code ${\bf C}$ contains $t$ codewords with the weight $w$ satisfying $tw=n$ and disjoint supports, then a nonzero retrieval rate star product PIR scheme can not protect against more than $w-1$ colluding servers no matter how to choose the retrieval code.}\\

It is obvious that there are many presently known good linear codes with the small defects satisfy the condition in Theorem 5.1. Therefore the upper bound on the privacy number of star product PIR schemes with colluding servers in Theorem 5.1 can be thought as an essential nature of star product PIR schemes. For star product PIR scheme with colluding servers for Reed-Muller $RM(m, r)$-coded distributed storage system, it can not protect against more than $2^{m-r}-1$ colluding servers no matter how to choose the retrieval code.\\

\section{Cyclic code based star product PIR scheme with colluding servers}

We prove the following result. This is a direct generalization of Theorem 1 in \cite{Cascudo} and was first proved in \cite{BMR21}.\\

{\bf Theorem 6.1.} {\em Let $q$ be a prime power and $n$ be a positive integer satisfying $\gcd(n, q)=1$. Let ${\bf C}$ and ${\bf D}$ be two length $n$ cyclic codes with the generator polynomials $${\bf g}_1(x)=\frac{x^n-1}{\prod_{i \in S_1} (x-\beta^i)},$$ and $${\bf g}_2(x)=\frac{x^n-1}{\prod_{i \in S_2} (x-\beta^i)},$$ where $\beta$ is a primitive $n$-th root of unity, $S_1$ and $S_2$ are the union of disjoint cyclotomic cosets. Then the star product ${\bf C} \star {\bf D}$ is a cyclic code with the generator polynomial $${\bf g}=\frac{x^n-1}{\prod_{i\in S_1+S_2} (x-\beta^i)}.$$}\\

{\bf Proof.} Let us recall some basic fact about cyclic codes. Let $\beta \in {\bf F}_{q^m}$ be the primitive $n$-th root of unity. For a subset ${\bf S} \subset {\bf Z}/n{\bf Z}=\{0, 1, \ldots, n-1\}$, let $B(S) \subset {\bf F}_{q^m}^n$ be the linear code $$B(S)=\{(f(\beta^0), f(\beta), \ldots, f(\beta^{n-1})): f=\Sigma_{i \in S} a_i x^i, a_i \in {\bf F}_{q^m}\}.$$ Then ${\bf C}=B(S)|_{{\bf F}_q}$, see \cite{Bier}. We have ${\bf C} \star {\bf D}=(B(S_1) \otimes B(S_2))|_{{\bf F}_q}$, actually $B(S_1) \otimes B(S_2)=\{(G(\beta^0), G(\beta^1), \ldots, G(\beta^{n-1}): G(x)=(\Sigma_{i \in S_1}  a_i x^i) \cdot (\Sigma_{j \in S_2} b_j x^j), a_i \in {\bf F}_{q^m}, b_j \in {\bf F}_{q^m}\}$. Then $$B(S_1) \otimes B(S_2)=\{(G(\beta^0), G(\beta^1), \ldots, G(\beta^{n-1})): G(x)=\Sigma_{i \in S_1+S_2} a_i x^i, a_i \in {\bf F}_{q^m}\}.$$ The conclusion follows immediately.\\

\subsection{BCH code based star product PIR schemes}

In this subsection let the storage code be an BCH code ${\bf C}_{q, n, b_1, \delta_1}$ and the the retrieval code ${\bf D}$ be the dual of an BCH code $({\bf C}_{q, n, b_2, \delta_2})^{\perp}$. Then the star product PIR scheme protects against any $\delta_2-1$ colluding servers. Set $S_1=(\bigcup_{0 \leq i \leq \delta_1-2} C_{b_1+i})^c$ and $S_2=\bigcup_{0 \leq j \leq \delta_2-1} C_{-b_2-j}$. Then the generator polynomial of ${\bf C} \star {\bf D}$ is $$\frac{x^n-1}{\prod_{i \in S_1+S_2} (x-\beta^i)}, $$ where $A^c={\bf Z}/n{\bf Z}\setminus A$. Because cyclic codes are transitive the retrieval rate of this star product PIR scheme is $$\frac{n-|S_1+S_2|}{n}.$$ Therefore  we need to find suitable $b_1, b_2$  such that the number of elements in $S_1+S_2$ is as small as possible. Let $N$ be the number of all cyclotomic cosets. Then $$|S_1+S_2| \leq (N-\delta_1+1)(\delta_2-1)m^2.$$ We have the following result.\\

{\bf Proposition 6.1.} {\em Let $q$ be a prime power and $n$ be a positive integer satisfying $\gcd(n, q)=1$. Let $N$ be the number of all cyclotomic cosets. Then we have a star product PIR scheme with the following parameters, the storage rate $$R_{storage} \geq \frac{n-(\delta_1-1)m}{n},$$ the ratio of tolerated failed servers $$f=\frac{\delta_1-1}{n},$$ the ratio of colluding servers $$t=\frac{\delta_2-1}{n}$$ and the retrieval rate $$R_{retrieval} \geq \frac{n-(N-\delta_1+1)(\delta_2-1)m^2}{n}.$$}\\

{\bf Example 6.1.} Let $q$ be a prime power and $n=q-1$. Then each cyclotomic coset has one element and there are $n$ cyclotomic cosets. $R_{storage}=\frac{n-\delta_1+1}{n}$, $f=\frac{\delta_1-1}{n}$, $t=\frac{\delta_2-1}{n}$. In this case if $b_1=n-1-\delta_2+2, b_2=1$, then $S_1+S_2=\{-\delta_2+1, \ldots, 0, 1, \delta_1-1\}$. Then the retrieval rate is $$R_{retrieval}=\frac{\delta_2-\delta_1+2}{n}.$$ This is the MDS code based star product PIR scheme with colluding servers attaining the Singleton type upper bound in \cite{Chen}.\\

{\bf Example 6.2.} Let $q$ be a prime power and $n=q^2-1$. There are $q-1$ cyclotomic cosets with one element of the form $C_{j(q+1)}$ where $j=0, 1, \ldots, q-2$. The other $\frac{(q-1)(q-2)}{2}$ cyclotomic cosets have two elements. Totally there are $N=\frac{q(q-1)}{2}$ cyclotomic cosets. If $\delta_2=5$, then we have a cyclic code based PIR scheme protects against $4$ colluding servers, and with the storage rate at least $$R_{storage}=1-\frac{\delta_1-1}{q^2-1}$$ and the retrieval rate at least  $$1-\frac{4(2q^2-2q-4\delta_1+4)}{q^2-1}.$$

If the number of tolerated failed servers is not cared and only the storage rate is important, we can take the disjoint union of several cyclotomic cosets as the set $S_1$. \\

{\bf Theorem 6.2.} {\em Let $q$ be a prime power and $n$ be a positive integer satisfying $\gcd(n, q)=1$. Let $S_1$ be the disjoint union of several cyclotomic cosets and $\delta_2$ be a positive integers. Then we have an cyclic code based star product PIR scheme with the following parameters, the storage rate $$R_{storage} = \frac{|S_1|}{n},$$ the ratio of colluding servers $$t=\frac{\delta_2-1}{n}$$ and the retrieval rate $$R_{retrieval} \geq \frac{n-m|S_1|(\delta_2-1)}{n}.$$}\\

\subsection{Low cost cyclic code based star product PIR schemes with colluding servers}

When the maximal weight $M$ of codewords in a storage code is much smaller than the number $n$ of servers. The computation cost and communication cost in the star product PIR scheme with colluding servers, are upper bounded by a $M$ factor.  Therefore the costs of computation and communication in the cyclic code based star product PIR scheme for this kind of special storage code are low. Let $m$ is an even positive integer and $e$ be a positive integer satisfying $e| 2^{\frac{m}{2}}+1$. Let $\alpha \in {\bf F}_{2^m}$ be a primitive element, $m_e(x)$ be the minimal polynomial of $\alpha^{e}$. Then $m_e(x)|x^n-1$, where $n=\frac{2^m-1}{e}$. Let ${\bf C}_e$ be the length $\frac{2^m-1}{e}$ cyclic code with the generator polynomial $m_e(x)$ and ${\bf C}_e^{\perp}$ be its dual. These kinds of irreducible cyclic code and their duals have been studied in the classical paper \cite{BM72}. The dual code ${\bf C}_e^{\perp}$ is a binary linear $[n, m, \frac{2^{m-1}-(e-1)2^{\frac{m}{2}-1}}{e}]_2$ code. It has only two nonzero weights with the weight distribution as follows, $$A(z)=1+\frac{2^m-1}{e}z^{\frac{2^{m-1}-(e-1)2^{\frac{m}{2}-1}}{e}}+\frac{(e-1)(2^m-1)}{e}z^{\frac{2^{m-1}+2^{\frac{m}{2}-1}}{e}}.$$ The maximal weight is $$M=\frac{2^{m-1}+2^{\frac{m}{2}-1}}{e}$$ is much smaller than the number $n$ of servers. Hence when the storage code is ${\bf C}^{\perp}$, the communication and computation costs of the star product PIR scheme are low. \\

Now we construct the retrieval code. As in the previous subsection we need find an BCH code ${\bf C}_{q, n, b, \delta}$ such that the set $S_1+S_2$ have as few elements as possible, where $$S_1=\{1, 2, \ldots, 2^{m-1}\},$$ and $$S_2=\bigcup_{0 \leq j \leq \delta-1} C_{-b-j}.$$ We have the following result.\\

{\bf Theorem 6.3.} {\em  Let $m$ be an even positive integer, $e|2^m-1$ be a positive divisor, $n=\frac{2^m-1}{e}$ and $\delta\leq 2^t+1$ where $t$ is a positive integer. $W$ files in ${\bf F}_2^{mb}$ are stored in the ${\bf C}_e^{\perp}$-coded distributed storage system. Then we have a low cost cyclic code based star product PIR scheme with the following parameters, the storage rate $$R_{storage} = \frac{m}{n},$$ the ratio of colluding servers $$t=\frac{\delta-1}{n}$$ and the retrieval rate $$R_{retrieval} \geq \frac{n-(m-t)2^{t+1}}{n}.$$ The computation cost and the communication cost are upper bounded by $(\frac{2^{m-1}+2^{\frac{m}{2}-1}}{e})Wb$ bit operations.}\\

{\bf Proof}. Set $b=0$, each element in $C_{-j}$, $0 \leq j \leq \delta-2 \leq 2^t-1$, has at most $t$ digits in its $2$-adic expansion. Then each element in $S_1+S_2$ has at most $t+1$ digits. The conclusion follows immediately.\\

{\bf Example 6.3.} Set $m=10$, $e=3$ and $\delta=17$. $W$ files in ${\bf F}_2^{mb}$ are stored in the distributed storage system. Then the distributed storage system has $n=341$ servers. The storage rate is $\frac{10}{341}$ and the ratio of tolerated failed servers is $f=\frac{159}{341}$. The star product PIR scheme protects against $16$ colluding servers. The retrieval rate is at least $R_{retrieval} \geq \frac{149}{341}$. The computation cost and the communication costs are upper bounded by $176Wb$ bit operations.

\section{Extended cyclic retrieval codes in PIR schemes for RM-coded distributed storage systems}

 For a linear code ${\bf E} \subset {\bf F}_2^{2^m-1}$, the extended code of ${\bf E}$ is $$Ext({\bf E})=\{(\Sigma_{i=1}^{2^m-1} x_i,  x_1, \ldots, x_{2^m-1}): (x_1, \ldots, x_{2^m-1}) \in {\bf E}\}.$$ In this section Reed-Muller codes are considered as extended cyclic codes, see \cite{BS17,DLX}. The storage code is $RM(m, r)=Ext({\bf C})$ where ${\bf C} \subset {\bf F}_2^{2^m-1}$ is a cyclic code of distance $2^{m-r}-1$. Let $\alpha \in {\bf F}_{2^m}$ be a primitive element. Cyclotomic cosets module $n=2^m-1$ are considered. For each $a \in {\bf Z}/n{\bf Z}$, the $2$-weight of $a$ is the number of nonzero digits in its $2$-adic expansion $$a=a_t 2^t+\cdots+a_12+a_0,$$ where $a_i \in \{0, 1\}$. Then the defining set of the binary cyclic code ${\bf C}$ is the disjoint union of cyclotomic cosets $$\bigcup_{wt(a) \leq m-r} C_a.$$ It was proved in \cite{Assmus} that $Ext({\bf C})$ is $RM(m, r)$. The retrieval code is the dual code ${\bf D}=Ext({\bf C}_{2, 2^m-1, 0, 2^{r'+1}-1})^{\perp}$ of the extended code of an BCH code ${\bf C}_{2, 2^m-1, 0, 2^{r'+1}-1}$ of the designed distance $2^{r'+1}-1$. Though they are not transitive code we prove that the retrieval rate is $\frac{\dim(({\bf C} \star {\bf D})^{\perp})}{2^m}$.\\

{\bf Proposition 7.1.} {\em $Ext({\bf C}_{2, 2^m-1, 0, 2^{r'+1}-1})^{\perp}=(0, {\bf C}_{2, 2^m-1, 0, 2^{r'+1}-1}^{\perp}) \oplus {\bf 1}$, where $$(0, {\bf C}_{2, 2^m-1, 0, 2^{r'+1}-1})=\{(0, {\bf c}): {\bf c} \in {\bf C}_{2, 2^m-1, 0, 2^{r'+1}-1}\},$$ and ${\bf 1}$ is the all $1$ vector.}\\

{\bf Proof.} It is clear that $$(0, {\bf C}_{2, 2^m-1, 0, 2^{r'+1}-1}^{\perp}) \oplus {\bf 1} \subset Ext({\bf C}_{2, 2^m-1, 0, 2^{r'+1}-1})^{\perp}.$$ Since the vector ${\bf 1}$ is not in $(0, {\bf C}_{2, 2^m-1, 0, 2^{r'+1}-1}^{\perp})$, the dimension of the left side is $2^m-1-\dim({\bf C}_{2, 2^m-1, 0, 2^{r'+1}-1})+1$. The dimension of the right side is $2^m-\dim({\bf C}_{2, 2^m-1, 0, 2^{r'+1}-1})$. \\

From Proposition 7.1 $Ext({\bf C}) \star {\bf D}=(0, {\bf C} \star {\bf C}_{2, 2^m-1, 0, 2^{r'+1}-1}^{\perp})\oplus RM(m, r)$. Then $$(Ext({\bf C} \star {\bf D}))^{\perp} \subset (*, ({\bf C} \star {\bf C}_{2, 2^m-1, 0, 2^{r'+1}-1}^{\perp})^{\perp}) \bigcap RM(m, m-r-1),$$ where  $$(*, {\bf E})=\{(x, {\bf c}): {\bf c} \in {\bf E}, \forall x\in {\bf F}_2\}.$$

Let ${\bf G}$ be the order $n=2^m-1$ cyclic subgroup of ${\bf S}_{n+1}$ generated by the element which keeps the zero position invariant and shifts on $\{1, \ldots, 2^m-1\}$ positions. Then ${\bf G}$ is a subgroup of the affine group of ${\bf F}_2^m$. By using this group as $G$ and $H$ in Lemma 2, page 2113 of \cite{Freij1}, we generate these subsets in Theorem 3. The zero position is in the information sets claimed in Theorem 3 of \cite{Freij1}. The number of information sets of $Ext({\bf C})$ and $Ext({\bf C}) \star {\bf D})^{\perp}$ containing this zero position is exactly $2^m-1$. The we prove the following result.\\

{\bf Theorem 7.1.} {\em The retrieval rate of the above star product PIR scheme with colluding servers is $\frac{2^m-1-\dim ({\bf C} \star {\bf C}_{2, 2^m-1, 0, 2^{r'+1}-1}^{\perp})}{2^m}$.}\\

It is clear $\dim({\bf C}_{2, 2^m-1, 0, 2^{r'+1}-1})>\dim(RM(m, m-r'-1))$ and $RM(m, m-r'-1) \subset {\bf C}_{2, 2^m-1, 0, 2^{r'+1}-1}$, because the defining set of an BCH code is smaller than the defining set of the punctured RM code. The retrieval rate in the above PIR scheme is bigger than $$\frac{\Sigma_{i=0}^{m-r-r'-1} \displaystyle{m \choose i}}{2^m}.$$ When $r$ and $r'$ is relatively big, the difference is big. The above result gives another interpretation of the comparison result in Section V of \cite{BMR21}.\\

{\bf Example 7.1.} Set $r=1$. The storage code $RM(m, 1)$ is the the extended code of cyclic code with the defining set $$\bigcup_{wt(a) \leq m-1} C_a.$$ The set $S_1=\{2^{m-1}+2^{m-2}+\cdots+2^{h+1}+2^{h-1}+\cdots+1: h=1, 2, \ldots, m-1\} \cup \{2^{m-2}+\cdots+2+1\}$ has $m$ elements. We set $r'=2$, the PIR scheme protects against $8-1=7$ colluding servers. Then the set $S_2=C_0 \bigcup C_{-1} \bigcup C_{-3} \bigcup C_{-5}$. Now the number of elements in the set $S_1+S_2$ needs to be calculated. There are $m+\frac{m(m-1)}{2}$ elements in the set $S_1+(C_0\bigcup C_{-1})$. On the other hand each element in $C_{-3}$ and $C_{-5}$ has its $2$-weight $2$. Each element in $C_{-3}$ is of the form $-2^t-2^{t-1}$ or $-2^{m-1}-1$, and each element in $C_{-5}$ is of the form $-2^t-2^{t-2}$, or $-2^{m-2}-1$ or $-2^{m-1}-2$. Therefore there are at most $$m+\frac{m(m-1)}{2}+2m^2$$ elements in $S_1+S_2$. The retrieval rate is at least $$\frac{2^m-1-m-\frac{m(m-1)}{2}-2m^2}{2^m},$$ when the retrieval code is the dual of an extended BCH code. This is bigger than the retrieval rate $$\frac{2^m-1-m-\frac{m(m-1)}{2}-\frac{m(m-1)(m-2)}{6}}{2^m},$$ when the retrieval code is $RM(m, 2)$ ($d(RM(m, 2)^{\perp})=8$). \\

\section{Conclusion}

In the design of star product PIR schemes with colluding servers, to find a good linear retrieval code for an efficient storage code is an interesting problem for coding theory. First of all we proved that when the storage code contains some special codewords, star product PIR schemes can not protect more than a small number of colluding servers. We gave examples to show that when the storage code is cyclic, the best choice of the retrieval code is not cyclic code in general. On the other hand it was showed that for the Reed-Muller coded distributed storage system, the best choice of the retrieval code is not the Reed-Muller code as used in previous papers. We also gave the parameters of cyclic code based star product PIR schemes with colluding servers. From results in this paper the star product PIR schemes with colluding servers with the storage and retrieval codes in the same family of algebraic codes seem not always efficient.\\

{\bf Acknowledgement.} We thank Dr. Jie Li for helpful discussion to improve the presentation of this paper. We thank Professor D. Ruano for providing us the 2nd version of the paper \cite{BMR21}.\\

\end{document}